\documentstyle[epsf]{mn}

\def\plotone#1{\centering \leavevmode
\epsfxsize=9.0truecm \epsfbox{#1}}

\begin{document}
\title[X--ray spectrum of BY Cam]
{Complex Absorption and Reflection of a Multi--temperature
Cyclotron--Bremsstrahlung X--ray Cooling Shock in BY Cam}
\author[C. Done and P. Magdziarz]
{C. Done$^1$ and P. Magdziarz$^{2,3}$\\
$^1$Department of Physics, University of Durham, South Road, Durham, DH1 3LE\\ 
$^2$Astronomical Observatory, Jagiellonian University, Orla~171, 
30-244 Cracow, Poland\\
$^3$N. Copernicus Astronomical Center, Bartycka 18, 00-716, Warsaw, Poland\\
}

\maketitle

\begin{abstract}

We re--analyse the ASCA and GINGA X--ray data from BY Cam, a slightly
asynchronous magnetic accreting white dwarf. The spectra are strongly affected
by complex absorption, which we model as a continuous (power law) distribution
of covering fraction and column of neutral material. This absorption causes a
smooth hardening of the spectrum below $\sim 3$ keV, and is probably produced by
material in the preshock column which overlies the X--ray emission region. The
ASCA data show that the intrinsic emission from the shock is not consistent with
a single temperature plasma. Significant iron L emission co--existing with iron
K shell lines from H and He--like iron clearly shows that there is a wide range
of temperatures present, as expected from a cooling shock structure.  The GINGA
data give the best constraints on the maximum temperature emission in the
shocked plasma, with $kT_{\rm max}=21^{+18}_{-4}$ keV. Cyclotron cooling should
also be important, which supresses the highest temperature bremsstrahlung
components, so the X--ray data only give a lower limit on the mass of the white
dwarf of $M\ge 0.5M_\odot$.  Reflection of the multi--temperature bremsstrahlung
emission from the white dwarf surface is also significantly detected.

We stress the importance of modelling {\it all} these effects in order to gain a
physically self--consistent picture of the X--ray spectra from polars in general
and BY Cam in particular. 

\end{abstract}
 
 
\section{INTRODUCTION}

Magnetic cataclysmic variables (polars or AM Her stars) are binary
systems where a magnetised ($B\sim 10^7$ G) white dwarf accretes from
a low mass companion (see e.g. the review by Cropper 1990). Such
magnetic fields are strong enough to disrupt disk formation. Instead,
the accreting stream is entrained by the magnetic field and falls
freely through the gravitational potential until it hits the white
dwarf surface. The resultant strong shock has a typical temperature of
tens of keV for optically thin material, giving rise to an X--ray emitting
plasma. Normally such systems are locked into synchronous rotation by
the magnetic field, so that (to zeroth order) the stream impacts onto
the same part of the magnetosphere. However, there are now three
systems known where the orbital period and white dwarf spin period are
slightly different, BY Cam, V1500 Cyg and RXJ 1940.2-1025 (e.g. Watson et al
1995), so that the accretion geometry changes continuously.

Recent progress in understanding the detailed shape of the X--ray
emission has concentrated on the effects of reprocessing of the hard
X--ray emission. Inclusion of the effect of Compton reflection from
the white dwarf surface {\it and} complex absorption from the
accreting material {\it and} the multi--temperature shock structure
are necessary ingredients in the spectral model. While the importance
of any one or two of these individual effects have long been
recognised (e.g. Imamura \& Durisen 1983 -- absorption and shock
structure; Swank, Fabian \& Ross 1984; Beardmore et al 1995 --
absorption and reflection) all three are necessary in order to obtain
a self--consistent description of the observed X--ray emission (Done
et al 1995; Cropper et al 1997). Both the complex absorption and
Compton reflection harden the spectrum, leading to an overestimate of
the plasma temperature. The line emission from the hot plasma gives
another independent measure of the plasma temperature, from the ratio
of 6.7 and 6.9 keV lines from He-- and H--like iron, respectively.
The temperature of the line emitting material is then lower than that
inferred from the 'observed' continuum. This discrepancy is exacerbated
in detectors of low energy resolution as Compton reflection produces a
fluorescence iron line at 6.4 keV from the (nearly neutral) X--ray
illuminated white dwarf surface, which is then blended by the
instrument with the 6.7 and 6.9 keV lines. This reduces the mean iron
line energy and leads to an underestimate of the plasma
temperature. This systematic mismatch between `observed' line and
continuum temperatures is particularly evident in the large sample of
GINGA (low spectral resolution) X--ray spectra of polars (Ishida
1991), even after the inclusion of complex absorption. Both complex
absorption {\it and} Compton reflection are required to obtain
consistent temperatures from the line and continuum data (Beardmore et
al 1995; Done et al 1995), but then the derived temperature is of
order 10 keV, much lower than the expected Rankine--Hugoniot temperature
for a strong shock of $\sim 57 (M/M_\odot) (R/5.57\times 10^8)^{-1}$ keV
(e.g. Frank, King \& Raine 1992, Imamura et al 1997). Using
multi--temperature shock models rather than assuming a single
temperature plasma goes some way to resolving this, 
as the mean temperature of a
$\sim 30$ keV multi--temperature shock structure is $\sim 10$ keV
(Imamura \& Durisen 1983; Done et al 1995; Wu, Chanagen \& Shaviv 1995;
Cropper et al 1997).
Cyclotron cooling of the hottest/least dense material probably accounts
for any remaining temperature discrepancy (Wu et al., 1995; 
Woelk \& Beuermann 1996).

While much real progress has been made, problems still remain.  The
most pressing of these have to do with the pre--shock accretion
column. This material is irradiated by the X--ray emission, so should
have a complex ionization structure (see e.g. Ross \& Fabian 1980;
Swank et al 1984; Kallman
et al 1993, 1996). It is also spatially extended over the X--ray source so
that different segments of the X--ray emission travel through
different path lengths of the material (Done et al 1995), especially
as the accretion column is probably arc--like in cross-section, rather
than circular (e.g. Cropper 1990). The column should also
add further complexity from secondary emission (Kallman et al 1993,
1996) and scattering (Ishida et al 1997).
Models including all these
effects are currently being developed (Rainger et al 1997, in preparation), but the
situation is clearly complex. Even more degrees of freedom are
possible in the models: the material in the column is probably
highly inhomogeneous, as dense 'ribbons' of material embedded in a
much less dense medium are required in order to explain the soft
X--ray excess seen in several of these objects (e.g. Kuijpers \& Pringle 1982; 
Frank, King \& Lasota 1988; Ramsay et al., 1994). 

BY Cam has been extensively studied by previous X--ray instruments, and its
spectrum shows many of the temperature discrepancies discussed above.  Kallman
et al (1993, 1996) propose an explanation for these in terms of secondary
emission from the column, producing an additional source of ionised 6.7 keV iron
line emission. Here we reanalyse the GINGA and ASCA data from BY Cam and show that
the spectrum is well described by the absorption/reflection/multi--temperature
shock picture developed above, and that the contribution of the pre--shock
material to the iron K line emission is probably negligible.

\section{OBSERVATIONS}

The GINGA data were taken in 1988 February 7--10, with background
observations taken directly before and after this. However, the second
background could not be used as the ROSAT All Sky Survey Bright Source
Catalog gives 3 sources within 35 arcmin of the pointing position
with total flux 1/5 of that from BY Cam itself. The first background
alone is not sufficiently long to sample all the background conditions
seen during the source observation so we exclude those parts of the
source data which are not well matched by the background observation.
Standard selection and cleaning criteria were then applied to both
source and background data, and the data were attitude corrected after
a 'local' background subtraction (Hayashida et al 1989; Williams et al
1992), resulting in 46 ks of data. 

Ishida et al (1991) describes the
time variability of the source during these observations. Clear
(though generally partial) eclipses are seen in the first half of the dataset,
but are absent in the second half. This is explained as a transition
in the accretion geometry from a two pole state, where only one of the accretion
regions is
periodically eclipsed by the white dwarf surface, to a state where the accretion
only takes place onto the non--eclipsing pole
(see also Silber et al 1992). We split the data into 3
spectra as in Ishida et al (1991), namely pulse high and pulse low
(double pole accretion with the eclipsing pole oriented towards and away from the
observer respectively) and flaring (accretion onto a single 
non-eclipsing pole).

The ASCA data taken on 1994 March 10-11 observations and were
previously reported by Kallman et al.\ (1996).  FTOOLS 3.6 was used to
generate standard response and arf files.  The SIS event files were
screened in bright2 mode with 1024-binning.  The GIS3 data had to be
binned onto 128 channels as the data were taken during the time when
the onboard CPU mode was wrongly set. This resulted in a total
exposure time of 22000 seconds and total count rate of 3.4 counts
per second over the four detectors after the standard reduction. The
data were accumulated into one single spectrum as Kallman et al (1996) 
show no obvious hardness ratio changes. 
We restricted the energy range to 0.6-10 keV for SIS data and 0.8-10 keV
for GIS data, and rebinned to have at least 20 counts per bin as 
required for validity of the $\chi^2$ statistic. 
The relative normalization of the detectors was within 10 \% from 
our preliminary analysis, thus we fitted all the spectra as a single 
data group. 

Kallman et al (1996) show the data folded on the polarization period 
of Mason, Liebert \& Schmidt (1989) with arbitrary phase.
The orbital coverage is $\sim 85$\% complete, and shows no evidence for
an eclipse. The gaps in phase coverage are in two 
segments, each of width $\sim 0.07-0.08$ in phase.  
The GINGA eclipse in the pulse state has width
$\ge 0.1$ in phase (Ishida et al 1991), so it seems unlikely that
the ASCA data could have avoided the eclipse altogether. 
The lack of
any convincing ingress/egress in the ASCA data then support (although they
cannot confirm) the view that the accretion geometry in BY Cam at the time of
the ASCA observation was analogous to that of the GINGA flaring state data.

\section{SPECTRAL FITTING}

Version 9.1 of XSPEC (Arnaud 1996) is used to fit the spectra.  The {\tt mekal}
code is used to model a single temperature hot plasma, while a modified version
of the {\tt cevmkl} code (see Done \& Osborne 1997) is used for the
multi--temperature plasma models which uses Morrison \& McCammon (1983)
abundance ratios. This gives a spectrum 
$$ f(\nu)\propto \int_{10^{5.5} K}^{kT_{max}} {\epsilon(n,T,\nu)\over 
\epsilon(n,T)} \Bigl({T\over T_{max}}\Bigr)^{\alpha-1} dT $$
where $\epsilon(n,T,\nu)$ is the bremstrahlung spectral emissivity
at a given density $n$ and temperature $T$
while $\epsilon(n,T)$ is the total (frequency integrated) emissivity
(see also Done et al 1995). Models with $\alpha=1$ give a spectrum where each
temperature component is weighted by its cooling time, i.e. the spectrum
expected from pure bremsstrahlung cooling at constant pressure and gravitational
field (see the discussion in Done et al 1995 and references therein).

The reflection code is described in Done et al (1995, 1997),
and here we assume a fixed inclination angle to the reflector of
$60^\circ$. Error ranges are given as $\Delta\chi^2=2.7$ unless otherwise
stated, although this is an underestimate of the true 90\% confidence limits
where parameters are correlated.

\subsection{ASCA data}

Single temperature plasma models, where the abundances of all the
elements heavier than He are assumed to scale together, give fairly
good fit, with $\chi^2_\nu=1458/1272$, but the minimum temperature of
$kT=200_{-35}^{+170}$ keV shows that the continuum is much harder than can
plausibly be expected for optically thin accretion onto a white dwarf.
Inclusion of an $\alpha=1$ multi--temperature continuum model
does not
significantly improve the fit ($\chi^2_\nu=1455/1272$), and the
derived (maximum) temperature is even higher, with $kT_{max}=940^{+\infty}_{-310}$ keV,
as the multi--temperature spectrum includes the softer cooling
components. Clearly there is some distortion present that is hardening
the observed spectrum in the ASCA bandpass. Absorption is the most
obvious way to do this since Compton reflection only contributes
significantly to the spectrum above 5 keV, so is unlikely to strongly 
affect the ASCA data.  Simple absorption
(complete covering by neutral material) is already included in the
fit, so we try partial covering by neutral material.
Ishida (1991) shows that
such complex absorption is not uncommon in polars, and uses this model
to fit the GINGA data of BY Cam (Ishida et al., 1991). 
This gives a significantly better fit for the multi--temperature $\alpha=1$
plasma model, with $\chi^2_\nu=1414/1270$ for $N_H=2.3^{+1.2}_{-0.8}\times 
10^{22}$ cm$^{-2}$
and $C_f=0.20\pm 0.05$
but still results in a physically unreasonable maximum shock temperature
of $kT=165_{-55}^{+100}$ keV. A more realistic shock temperature of $kT=63^{+17}_{-15}$ keV 
can be obtained with the single temperature plasma model with
$N_H=2.7_{-1.0}^{+1.8}\times 10^{22}$ cm$^{-2}$ and $C_f=0.15\pm 0.03$, but at the
expense of physical consistency (we expect multi--temperature components to be
present) and goodness of fit ($\chi^2_\nu=1436/1270$).

The ASCA data extend to substantially lower energies than can be studied with
GINGA, and so are more sensitive to the form of the absorber. A partial covering
model is an {\it ad hoc} description of what is expected to be a much more
complex situation. A power law distribution
of column with covering fraction i.e. $C_f(N_H)\propto N_H^\beta$
(see also Norton, Watson \& King 1991)
gives a good approximation to the
complex absorption expected through neutral material extended over the
source (e.g. Done et al 1995). The sum of covering fractions
must be unity so that the emergent spectrum $S(E)$ is related to the
intrinsic spectrum $S_{\rm int}(E)$ by
\begin{eqnarray}
S(E) & = & S_{\rm int}(E) A 
\int_{N_{\rm H, min}}^{N_{\rm H, max}} N_H^\beta \exp (-N_H \sigma(E)) dN_H
\end{eqnarray}
where
\begin{eqnarray}
A & = & {\beta + 1\over N_{\rm H, max}^{\beta+1}-N_{\rm H, min}^{\beta+1}}
\ldots \beta \ne -1\nonumber \\
& = & {1\over \log (N_{\rm H, max}/N_{\rm H, min}) }
\ldots \beta = -1\nonumber\\
\nonumber\end{eqnarray}
and $\sigma(E)$ is the photo--electric absorption cross--section from Morrison
and McCammon (1983) abundances as given by the {\tt wabs} model in XSPEC. We fix
$N_{\rm H, min}=10^{20}$ cm$^{-2}$, and Figure 1 shows the 
transmitted fraction as a function of energy for $N_{H,max}=10^{24}$ cm$^{-2}$
with $\beta$ ranging from $-1$ to $1$. For comparison we also show
the transmitted fraction obtained by the circular column model of Done et al
(1995) for a mid column of $5\times 10^{23}$ cm$^{-2}$ (and hence a maximum column of
$10^{24}$ cm$^{-2}$), and that obtained from complete covering by a column of
$5\times 10^{23}$ cm$^{-2}$. The power law absorption model 
gives a smooth hardening of the
spectrum which matches the data significantly better ($\chi^2_\nu=1399/1270$)
than the partial covering model.  The derived maximum temperature of a
multi--temperature plasma is then the more physically reasonable $kT_{max}=45_{-14}^{+35}$
keV for a maximum column of $2.5^{+\infty}_{-1.8}\times 10^{24}$ cm$^{-2}$, with index $\beta =
-1.07\pm 0.08$.  Figure 2 shows the data and residuals for this model, along
with the intrinsic spectrum before the absorption.

\begin{figure}
\plotone{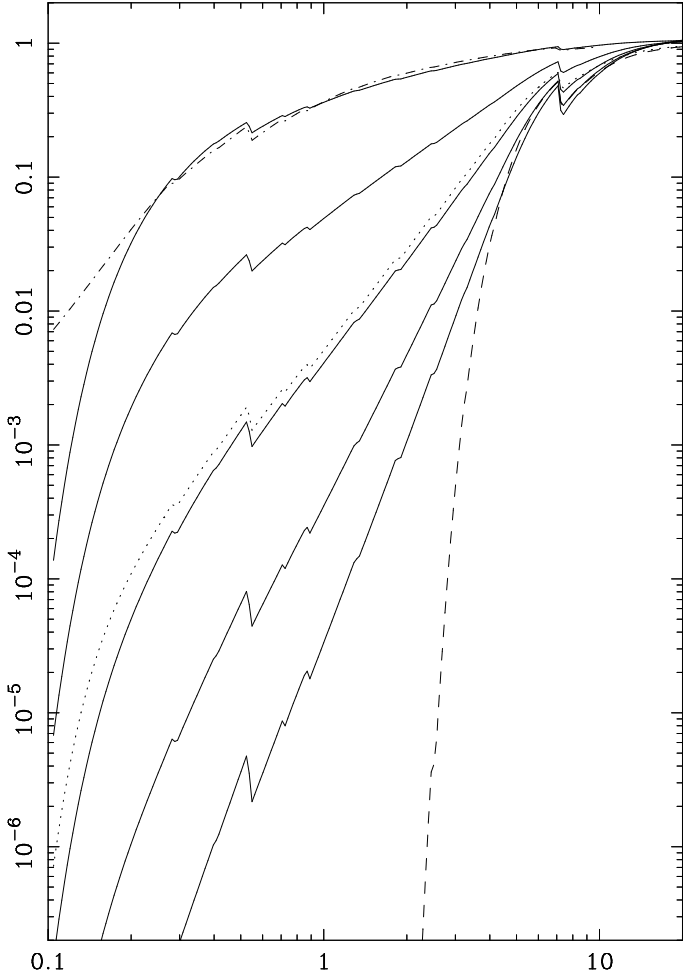}
\caption{
The transmission as a function of energy through neutral material with a power
law covering fraction -- column distribution for $\beta=-1,-0.5,0,0.5$ and $1$
from top to bottom. The dotted line shows the transmission through a circular
neutral column where $N_{H,mid}=5\times 10^{23}$ cm$^{-2}$ (Done et al 1995) 
while the dashed line shows the transmission through a complete screen with
$N_H=5\times 10^{23}$ cm$^{-2}$. The dash--dot line shows the transmission
through a circular column with radial density stratification (see Section 4.1).
}
\end{figure}

\begin{figure}
\plotone{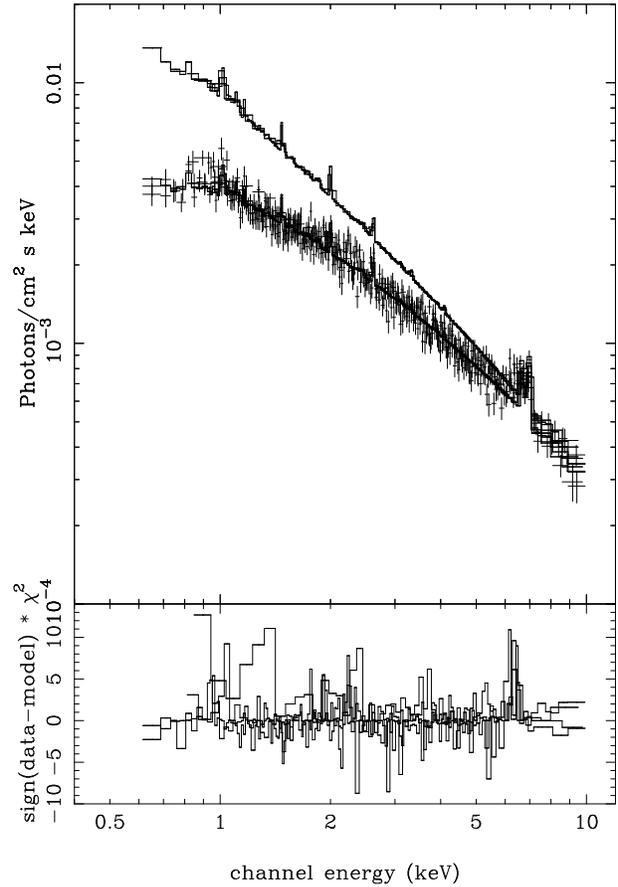}
\caption{
The ASCA spectrum of BY Cam, modelled by a multi--temperature $\alpha=1$ plasma,
absorbed by neutral material with a power law distribution of covering fraction
with column. The top panel shows this model and the unfolded spectrum, together
with the intrinsic emission before absorption, while the bottom panel shows the
remaining residuals to the fit. 
}
\end{figure}

Alternatively, the material may be ionized due to the intense irradiation of the
pre--shock column by the X--ray emitting plasma.  We use the XSPEC model {\tt
absori} to approximate this, fixing the power law temperature of the
illuminated material to $3\times 10^4$ K and approximating the X--ray spectrum by a
power law of photon index $\Gamma=1.5$.  Complete covering by a ionized material
gives a worse description of the data, with $\chi^2_\nu=1431/1270$ for a column
of $5.2^{+3.5}_{-2.5}\times 10^{21}$ cm$^{-2}$ at an ionization parameter of 
$\xi=240_{-120}^{+160}$, and
again with an unphysically high temperature for the $\alpha=1$ plasma of
$kT_{max}=270^{+130}_{-100}$ keV. Such models have been used above 2 keV to match the complex
absorption seen in the data (Beardmore et al 1995; Cropper et al 1997), but they
predict a large Oxygen edge which is inconsistent with the observed spectra
below 1 keV. 

To summarise the results so far, 
the data strongly require heavy absorption, which significantly
hardens the spectrum below 3 keV, but this absorption is complex.
Complete covering of the source by a screen of material in a single
ionization state (neutral or ionised) is strongly ruled out. Instead
the absorption can be described as a continuous range of columns
through which the source is seen, as expected for the pre-shock
material. While the ionization structure of
the column is outside of the scope of this paper (see Rainger et al
1997), the data are clearly consistent with a power law
column--covering fraction distribution of neutral material, so we use
this model in all the following fits. The continuum is also significantly better
described by a multi--temperature $\alpha=1$ cooling shock model than by a
single temperature plasma, so we use this 
to describe the plasma emission in all the following fits.

While the resultant fit is statistically adequate, there are clear
residuals left around the iron K line (see Figure 2).  
Adding a narrow Gaussian line
gives a reduction in $\chi^2_\nu$ to $1363/1268$ (significant at greater than
$99.9$ \% confidence) for $E=6.39_{-0.04}^{+0.03}$  keV
and equivalent width $86^{+36}_{-26}$ eV.
This additional line component is {\it inconsistent} with being at 6.7 keV,
as postulated by Kallman et al. (1996). A (nearly) neutral iron
fluorescence line is expected from reflection of the intrinsic
spectrum from the white dwarf surface. Fixing the line energy at 6.4 keV and 
including a neutral Compton reflection continuum
that must accompany any reflected line emission
leads to $\chi^2_\nu=1360/1268$, for an amount of reflection $R=1.5_{-1.3}^{+0.6}$
(where $R=1$ denotes the normalisation of the reflected continuuum expected from
an isotropically illuminated slab covering a solid angle of $2\pi$)
assuming the
abundances in the reflector are the same as those in the hot plasma,
with maximum temperature $kT_{max}=38^{+34}_{-13}$ keV. 
While the detection of the reflected continuum is only marginally significant, 
its level is consistent with that expected from the strength of the cold
iron fluorescence line. The line equivalent width and amount of 
reflection continuum are anti--correlated in the fitting, so a self
consistent reflection model (including both line and continuum) would
be much better constrained. 
This fit is detailed in Table 1, and shown in Figure 3.

\begin{figure}
\plotone{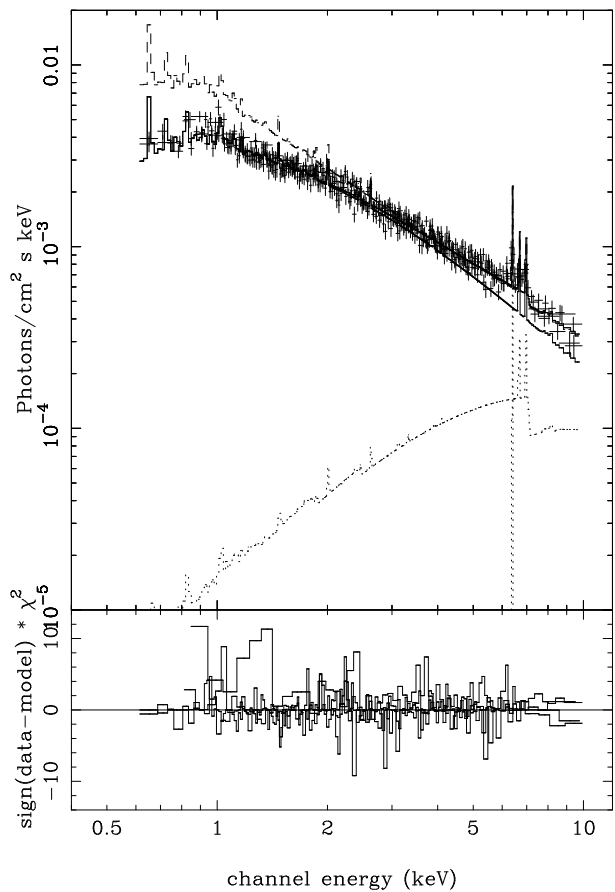}
\caption{
The ASCA spectrum of BY Cam, modelled by a multi--temperature $\alpha=1$ plasma
and its reflection from the white dwarf surface, 
absorbed by neutral material with a power law distribution of covering fraction
with column. The top panel shows this model and the unfolded spectrum, together
with the intrinsic emission before absorption, while the bottom panel shows the
remaining residuals to the fit.
}
\end{figure}

The abundances of the elements in the hot plasma need not all scale
together, especially as BY Cam is proposed to have abundance anomalies
to explain the enhanced NV line emission observed in its UV spectrum
(Bonnet--Bidaud \& Mouchet 1987). The N X--ray lines are below the 0.6
keV well calibrated low energy limit of the ASCA detectors, but can be 
investigated by including the 
less reliable data down to 0.4 keV. The soft excess is then also
apparent (Kallman et al 1996), so a blackbody is needed to model this.
The derived limit on the 
N abundance is then $\le 15\times $ solar, which is not restrictive.
Going back to the standard bandpass, the 
strongest lines expected at these plasma temperatures are Fe, Si, S and
O. Letting the abundances of all these be free gives 
$A_{Fe}=0.41_{-0.12}^{+0.43}$, 
$A_{Si}=0.69_{-0.31}^{+0.30}$, 
$A_{S}=0.39_{-0.39}^{+0.47}$,
$A_{O}=0^{+0.29}$ and 
$A_{rest}=0.38_{-0.38}^{+0.51}$
with $\chi^2_\nu=1353/1264$, i.e. there is no significant improvement
in the fit as the inclusion of 4 extra free parameters gives a reduction 
of only $\Delta\chi^2=7$. Thus there is no strong evidence for
non--solar abundance {\it ratios} of the elements, although O is
marginally lower than expected.

The intrinsic multi--temperature plasma emission can also be
investigated in more detail. We have so far assumed that the plasma
cools from the shock temperature to the white dwarf photosphere
at $\sim 10^{5.5}$ K $\approx 27$ eV, and
that all these cooler components contribute to the spectrum. However,
this is not necessarily the case since the cooling plasma may be dense
enough to become optically thick, so giving an apparent minimum
temperature to the cooling radiation which is rather higher than that
of the white dwarf. We modify the plasma emission model to include the
minimum temperature as a free parameter of the fit, and find that this
converges to an identical fit as before, with $kT_{min}=27$ eV.  The
limits on the O abundance are also unchanged, with $A_{O}=0^{+0.29}$
($\chi^2_\nu=1353/1266$), showing that incomplete cooling is not
responsible for any potential deficit of O line emission. 

Similarly, the temperature distribution need not be given by
the $\alpha=1$ power law model expected from pure X--ray line
and continuum cooling as the assumptions of bremsstrahlung only cooling at
constant pressure and
gravitational field may not be accurate. 
Letting $\alpha$ be free gives a 
significant decrease in $\chi^2_\nu$ to $1354/1267$ for $\alpha=0.6$
and $kT_{max}=55$ keV.

Thus the data are consistent with multi--temperature emission from
complete cooling behind the shock, where the accreting material is
$\sim 0.5\times$ solar abundance in all the elements.  This intrinsic spectrum
is then modified by reflection from the white dwarf surface, producing
both reflected continuum and associated 6.4 keV iron fluorescence line,
and further distorted by complex absorption.  
The intrinsic emission is required to be a
multi--temperature plasma, and the cooling components are
significantly detected in the data. 
Parameters for both the expected $\alpha=1$ cooling model and the
better fitting model where $\alpha$ is free 
are given in Table 1. The
best fit single temperature model (including reflection and complex
absorption) is also tabulated for comparison, but is a worse fit by
$\Delta\chi^2=20$ as the data contain significant iron L line
emission which cannot be fit by the single temperature models.  This
is the first observational confirmation of the theoretically expected
cooling of the shocked plasma in polars, although the distribution of
cool components with temperature is marginally inconsistent with the
predicted $\alpha=1$.

\begin{table*}
\begin{minipage}{180mm}
\caption{ASCA spectral fitting with multi--temperature (plT) and single
temperature (1T) models}
\label{}
\begin{tabular}{ccccccccccc}                \hline
 
Model &
$N_{H,Gal}$ & 
kT\footnote{This is the maximum temperature in the multi--temperature models} 
(keV) & A\footnote{with respect to Morrison and McCammon (1983)} & 
N\footnote{The normalisation of the plasma model in units of
$10^{-14}/(4\pi D^2)\times$ Emission Measure}$\times 10^{-2}$ & 
$\alpha$ & $R$ & EW (eV)
& $N_{\rm H,max}$\footnote{Units of $10^{22}$ cm$^{-2}$} & $\beta$ & 
$\chi^2_\nu$\\

\hline
plT & $0.07^{+0.03}_{-0.04}$ & $38^{+34}_{-13}$ & $0.52_{-0.15}^{+0.33}$ & 
$7.0$ &
$1$\footnote{Parameter fixed} & $1.5_{-1.3}^{+0.6}$ & $75\pm 25$ 
& $5.6_{-2.6}^{+\infty}$ & $-1.08^{+0.12}_{-0.13}$ & $1360/1268$ \\
plT & $0.09\pm 0.04$ & $55^{+\infty}_{-22}$ & $0.45^{+0.30}_{-0.20}$  & 
$5.1$ & 
$0.60\pm 0.25$ & $1.5^{+0.7}_{-0.6}$ & $75\pm 25$ & $4.6^{+7.0}_{-2.6}$ & 
$-1.0^{+0.25}_{-0.15}$ & $1354/1267$ \\
1T & $0.06\pm 0.03$ & $18_{-6}^{+5}$ & $0.49_{-0.18}^{+0.20}$ & 
$3.5$ &
$\infty^e$ & $1.1\pm 0.6$ & $85\pm 25$ & 
$1000_{-997}^{+\infty}$ & $-1.23\pm 0.09$ & $1380/1268$\\

\end{tabular}
\end{minipage}
\end{table*}

\subsection{GINGA data}

All 3 GINGA spectra give a lower flux than the ASCA data, even in the
overlapping 2--10 keV band.  Ratios of these spectra with the best fit ASCA
model are given in Figure 4. All the spectra are also systematically different
in shape compared to the ASCA data, being generally softer. Thus
there are significant spectral as well as intensity changes between the GINGA
and ASCA observations. 

\begin{figure}
\plotone{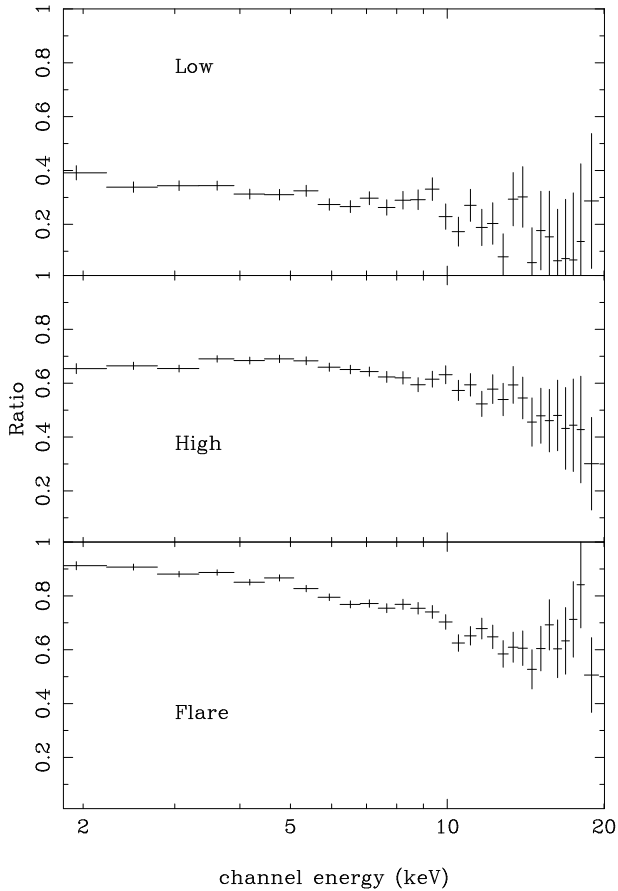}
\caption{
The ratio of the GINGA spectra to the best fit $\alpha=1$ ASCA spectral model
shown in Figure 3. The low pulse phase, high pulse phase and flare spectra are
shown in the top, middle and bottom panels, respectively. 
}
\end{figure}

We first fit the GINGA low and high pulse state data, where the 
inclination angle to the eclipsing spot
is changing with phase. We fit the two datasets simultaneously,
constraining the interstellar absorption, plasma temperature and abundances to
be the same. Ishida et al (1991) show that complex absorption is required by the
data, and model this by partial covering. A single temperature plasma model with
partial covering gives $\chi^2=37/51$. The low state spectrum is significantly
softer than the high state at low energies, and in fact does not require any
absorption. Constraining the two partial coverers to the be same gives a much
worse fit with $\chi^2=56/53$.  Replacing the partial coverer by the power law
column--covering fraction model used for the ASCA data gives an equally good
fit, showing that the wider ASCA bandpass is needed to properly constrain the
complex absorption properties. We use the power law absorber for consistency
with the ASCA results, but since $N_{\rm H, max}$ and $\beta$ are strongly
correlated and poorly determined we constrain $\beta$ to be equal across all the
datasets.  This gives $\chi^2_\nu=35/52$ for 
$\beta=-1.01\pm 0.08$, 
$N_{H,max}=5.0^{+9.0}_{-4.5}$ and $770^{+\infty}_{-690}\times 10^{22}$
cm$^{-2}$ for the low and high state spectra, respectively.

Again the residuals clearly indicate structure around the iron line and edge
energies so we fit a reflection continuum spectrum and 6.4 keV iron fluorescence
line. This gives a significant improvement in the fit with $\chi^2_\nu=28/50$.
The continuum was then replaced with the multi--temperature $\alpha=1$ emission
model. This did not give a significantly better fit to the data, showing that 
the 2--20 keV GINGA bandpass is not sufficient to distinguish between 
single and multi--temperature emission models. However, we chose this model 
for further study, for ease of comparison with the ASCA fits. The 
flaring state spectrum is then also fit with 
this model, and the results detailed in Table 2. Error ranges on the 
absorber are not given since this quantity is unconstrained by the
data. The harder spectrum seen in the high state can equally well be described
by a larger contribution from the reflection continuum as by more
absorption. 

\begin{table*}
\begin{minipage}{180mm}
\caption{Simultaneous spectral fits to the GINGA 
pulse low (Low), pulse high (High) and flaring data with a multi--temperature
plasma continuum, reflection and complex absorption. The temperature of the High
and Low pulse states are tied together, while the abundance,
temperature distribution and absorption power law are tied across
all the datasets.}
\label{}
\begin{tabular}{ccccccccccc}                \hline
 
Data &
$N_{H,Gal}$ & 
kT\footnote{This is the maximum temperature in the multi--temperature models} 
(keV) & A\footnote{with respect to Morrison and McCammon (1983)} & 
N\footnote{The normalisation of the plasma model in units of
$10^{-14}/(4\pi D^2)\times$ Emission Measure} $\times 10^{-2}$& $\alpha$ & $R$ & EW (eV)
& $N_{\rm H,max}$\footnote{Units of $10^{22}$ cm$^{-2}$} & $\beta$ & 
Total $\chi^2_\nu$\\
\hline
Low & $0.0^{+0.52}$ & $21_{-4}^{+18}$ & $0.25_{-0.09}^{+0.10}$ &
$2.4$ & $1$\footnote{Parameter fixed} & 
$1.1_{-1.1}^{+1.2}$ & $0^{+135}$ & $1.9_{-1.9}^{+\infty}$ & 
$-0.9$ & $41.4/73$\\
High &              &               &                    &
$5.2$ &
    & $1.6^{+1.2}_{-1.3}$ & $35^{+70}_{-35}$ & $8.8$ \\
Flaring &              & $25_{-7}^{+18}$          &                    &
$6.5$ &
    & $1.0_{-0.9}^{+1.0}$ & $45^{+80}_{-45}$ & $4.0$ \\

\end{tabular}
\end{minipage}
\end{table*}

\section{DISCUSSION}

\subsection{Absorption}

The complex absorption is clearly a major component in the spectrum, giving a
smooth hardening at energies lower than $\sim 4$ keV which cannot be described
by complete covering by material of a single ionization parameter.  The lack of
the standard absorption signatures (strong low energy cutoff or ionised edges)
means that it is not always obvious that this component is present in the
spectrum. In BY Cam its main signature is that the derived temperature
for the ASCA spectrum becomes unphysically high without it. 
We caution that where this complex absorption is present, the observed
bremsstrahlung luminosity can be severely underestimated in spectra which only
extend to 2--3 keV (e.g. ROSAT; Ramsay et al., 1994).

Our model for this absorption is one in which neutral material with a range of
covering fractions obscures the source, so that $C_f(N_{\rm H})\propto N_{\rm
H}^\beta$. The simplest geometry of a cylindrical neutral column overlaying a
circular source predicts a distribution of column with covering fraction such
that $\beta\sim 0$ (see Figure 1).  This is very different from the observed
$\beta=-1$, indicating that the situation is (unsurprisingly) 
much more complex.

The absorption can only be constrained by the ASCA data 
which is phase averaged,
so some of the discrepancy in expected column distribution could
arise from co--adding data. 
Further distortions could also be produced if the 
accretion column is ionised. This would have the 
effect of making the absorption from a circular column 
less dramatic than predicted from neutral
material, as required by the $\beta=-1$ distribution.
Photo--ionization of a uniform accretion column has long been known to 
be likely (Ross \& Fabian 1980; Swank, Fabian \& Ross 1984) but is much
less probable if the 
column has a highly inhomogeneous 'blobby'
structure (e.g. Frank, King \& Lasota 1988).
In this case radial structure can alter the observed properties.
The observed $\beta=-1$ distribution can be reproduced by a model
in which neutral blobs accrete preferentially
towards the centre of the column, with blob density $\propto r^{-5}$,
so that the much less of the X--ray source is covered 
by the highest column density material (dash--dotted line in Figure 1).

All accretion column models predict a substantial change in absorption with
phase.  For a neutral, constant density column, there should be much less
absorption when the hard X--ray emission region is on the limb of the white
dwarf (pulse low state) than when it is viewed more directly (pulse high) as the
mean path length through the accretion column is smaller (Done et al
1995). Allowing for ionization enhances this effect, as the high ionization
material (and hence weakest absorption) is produced in the layers closest to the
shock. The GINGA data are indeed consistent with a change in absorption between low and high
state, in the sense that the low state is less absorbed (see Table 2), but a complete model of
the density and ionization structure of the photo--ionised column to determine
both its absorption and emission properties is clearly needed before such ideas
can be properly tested.

The absorption is also presumably linked to the mass accretion rate, 
as there is more material in the column at high mass accretion rates.
Thus for a neutral column we expect that the absorption should correlate with
X--ray luminosity. By contrast, an ionised column may show absorption 
anti--correlated with luminosity as the bremsstrahlung intensity is 
$\propto n^2$ while the column density is only $\propto n$. Again,
proper modelling of the column is required in order to determine the
observational consequences of this, and better data are required to test 
whether the absorber is predominantly ionised or neutral.

\subsection{Continuum}

A single temperature plasma continuum is clearly ruled out by the ASCA
data. Multi--temperature plasma is expected physically, as the hot shocked gas
must cool as it emits X-ray radiation. The observational signature of this is
that the spectra show significant emission both from the H and He--like
K$\alpha$ lines at 6.9 and 6.7 keV and iron L shell lines around 1 keV from much
lower ionization species. For all of these different Fe ion states to co--exist
requires a multi--temperature plasma. The observed maximum X--ray temperature
of $21^{+18}_{-4}$ keV (from the better constrained GINGA fits) is 
rather lower than the $\sim 57$ keV expected from a pure bremsstrahlung 
shock above a solar mass white dwarf. Taken at face value this would limit the
mass of the white dwarf to $0.6^{+0.3}_{-0.1}$. However, BY Cam has a substantial magnetic field
of $\sim 28$ MG (Schwope 1996 revised from the Cropper et al 1989 value of 40.8
MG), so cyclotron cooling may be important. The high
temperature/low density material close to the shock preferentially cools via
cyclotron (emissivity $\propto T^{2.5}n^{0.15}$: Wu et al 1995) 
while at lower temperatures/higher densities
the bremsstrahlung (emissivity $\propto T^{1/2}n^2$) becomes more
important. This has the effect of reducing the maximum X--ray bremsstrahlung
temperature below that expected from a simple strong shock (Wu et al 1995;
Woelk \& Beuermann 1996).

Composite cyclotron--bremsstrahlung cooling shocks also have another feature not
seen in pure bremsstrahlung shocks, namely that the maximum X--ray temperature
{\it changes} as a function of mass accretion rate.  Higher mass accretion rates
increase the density of the shocked material, and so shift the balance between
cyclotron and bremsstrahlung cooling in favour of bremsstrahlung (Woelk \&
Beuermann 1996, Wu et al 1995).  Thus we expect the maximum shock temperature as
seen from the X--ray emission to increase with increasing mass accretion rate.
We use the code of Wu et al (1995) to calculate the temperature at which
bremsstrahlung and cyclotron cooling are equal for $B=30$ MG, $M=M_\odot$ and
mass accretion rate of $3.14$ and $6.28\times 10^{16}$ g
s$^{-1}$ (corresponding to luminosity of $0.75$ and $1.5\times 10^{34}$ ergs
s$^{-1}$, respectively). This gives $T_{\rm max}\sim 25$ and $35$ keV. These are
intruigingly close to the best fit derived maximum temperatures for the
$\alpha=1$ GINGA
and ASCA data of $20$ and $38$ keV, respectively, for a factor of $\sim 2$
difference in mean 2--10 keV count rate.  While the large error bars mean that the
effect is not significant here, these temperature changes should be seen in high quality, broad
bandpass spectra from SAX and/or XTE

Cyclotron cooling has a strong and variable effect in suppressing the high
temperature X--ray emission. These BY Cam data stress again the point made by Wu
et al (1995) and Woelk \& Beuermann (1996) that equating the maximum
bremsstrahlung temperature to the shock temperature is {\it not} a good guide to
the white dwarf mass. However, downstream from the shock,
once the density and temperature are such that
bremsstrahlung dominates the cooling then the shock structure is again given by
the standard bremsstrahlung $\alpha=1$ cooling. This can be seen from the
results of the code of Wu et al (1995) and also theoretically, since when
bremsstralung cooling dominates all the equations are the same as for a 
bremsstralung only shock.
The derived temperature distribution seen in the ASCA data 
is (marginally) significantly different from this, with $\alpha\sim 0.6$ i.e. more
cool components than expected. This may be an artifact of the
{\it ad hoc} absorption model used (see above). If the real absorber
includes complex contributions from ionised material then some structure should
be present at the OVIII K and Fe L line and edge energies 
which may distort the derived continuum.
Alternatively, this may be a real effect, showing that the shock structure
deviates significantly from that expected, perhaps due to conduction
and/or Compton cooling (Imamura \& Durisen 1983; Imamura et al 1987).

\subsection{Abundances}

The X--ray line emission from BY Cam is weaker than expected from solar
abundance (Morrison and McCammon 1983) coronal models. This has been seen
systematically from many classes of objects, including stellar coronae (see
e.g. Guedel et al 1997; Mewe et al 1997) as well as other Cataclysmic Variables
(Done et al 1995; Done \& Osborne 1997).
While deviations from coronal equilibrium may be expected, none of
these seem able to produce the observed weak lines (Done et al 1995,
Done and Osborne 1997), so it seems more likely that this represents a
true under-abundance with respect to solar. This seems more plausible
now that optical studies suggest stars in the solar
neighbourhood have considerable dispersion in metallicity, and that
the mean [Fe/H] is 0.25-0.3 dex {\it lower} than that of the Sun
(Edvardsson et al 1993), but is still an uncomfortable conclusion.

Strong UV lines from NV have led to a claim that the N abundance is
enhanced (Bonnet--Bidaud \& Mouchet 1987). X--ray line strengths from
H and He--like ions give much less ambiguous determinations of
abundances than UV or optical photo--ionised lines, but the X--ray N
lines are below the well calibrated energy limit of the detector and
the derived limit of $\le 15 \times$ solar is not restrictive.  Whether
or not N is indeed enhanced, the suggested explanation of a recent
novae explosion as the cause (Bonnet--Bidaud \& Mouchet 1987) seems
unlikely since the lines arise from the accretion stream i.e. material
from the {\it secondary} star. Even if the secondary did manage to
capture some fraction of the nuclear processed material, the elemental
abundance anomalies would become diluted by mixing with the
un--reprocessed material in the stellar atmosphere. However, recent
HST spectra of the surface layers of the accreting white dwarf in VW
Hydri also seem to show abundance anomalies, with Al$\sim 15\times$
solar, and P$\sim 900\times$ solar (Sion et al 1997)! Again, these
have been interpreted as being indicative of a recent novae, so we
examine other possible X--ray signatures of nuclear reprocessing in BY
Cam.

The core composition of the accreting white dwarf can either be CO or ONeMg.
For a CO white dwarf, calculations by Kovetz \& Prialnik (1997) show that while
N is almost always enhanced in a nova explosion, the O abundance can be almost
unaffected or strongly depleted, while the heavier elements are not
substantially enriched. This is consistent with the ASCA observations.
Conversely, for an ONeMg white dwarf core, N/O$\le 1$ requires that the
abundances of P and S should also be strongly enhanced (Politano et al
1995). This type of nova explosion can then be clearly ruled out since S is not
overabundant (see Section 3.1).  Thus if a recent novae has enriched the N
abundance, it would have to be from accretion onto a CO core, 
although a more likely explanation for anomalous C/N ratios is that the
secondary is evolved (Mouchet et al 1997).

\subsection{Comparison with previous work}

Physically motivated (as opposed to phenomolgical) descriptions of the spectrum
of BY Cam are given by Kallman et al (1996), using the same ASCA data as studied
here. However, their conclusions are rather different in that they propose a
strong {\it line emission} component at 6.7 keV from the pre--shock column,
rather than absorption. They require this because their assumed continuum 
form (a 30 keV bremsstrahlung) produces copious 6.9 keV line from H--like
iron, but has a low He--like iron ion fraction, so cannot produce much of the
observed 6.7 keV line. We show that this conclusion is obviated by using 
a continuum given by the (physically expected) multi--temperature plasma 
emission from a cooling shock, as this can produce both the 6.7 and 
6.9 keV lines. Kallman et al (1996) discount this possibility due to the
weakness of the observed 6.9 keV line compared to that expected from 
lower temperature plasma with solar abundances. This objection is removed by 
allowing iron (and the other element abundances) to be sub--solar. Kallman et al
(1996) also comment that with lower temperature plasma the observed continuum is
too flat. In our analysis this is resolved by the inclusion of complex
absorption from the column. We note that Kallman et al (1996) require a
pre--shock column of order $\sim 10^{23}$ cm$^{-2}$ overlaying the X--ray emission region.
Thus their model is not self--consistent as it does not take into account the
strong absorption that this predicts. Our model is also
inconsistent in not including the {\it emission} from the X--ray illuminated
column, but the sub--solar iron abundances means that any predicted iron K line
emission is reduced by a factor 3 below that of Kallman et al (1996) i.e. 
$\le 25$ eV equivalent width, which is smaller than the error bars. 
The photo--ionised column is instead expected to produce the 
bulk of its emission from continuum/recombination continuum/lines
at energies below $\sim 1$ keV, as noted by Kallman et al (1993) where a 
large neutral absorption column is required to suppress the strong predicted
low energy emission.

Kallman et al (1996) also speculate on the presence of a reflection component to
produce the observed 6.4 keV line. Their derived limit of $\Omega/2\pi\le 0.6$
is again assuming solar abundances, without including the inclination effects,
and using the results of early reflection calculations, which gave EW $\sim
210\Omega/2\pi$ eV for a 20 keV bremsstrahlung illumination. More recent Monte--Carlo
results (e.g. George \& Fabian 1991, Matt, Perola \& Piro 1991, Van Teesling,
Kaastra \& Heise 1996), 
scaled to a 20 keV bremsstrahlung
continuum (e.g. Beardmore et al 1995)
indicate that even for a face on, solar abundance slab 
the line emission is no more than $\sim
150 \Omega/2\pi$ eV, while for a mean (phase averaged) 
viewing angle of $60^\circ$ this reduces to $\sim 110 \Omega/2\pi$ eV, while
abundances of $\sim
0.4\times$ solar reduces it further to $\sim 90
\Omega/2\pi$ eV (George \& Fabian 1991). Thus the 6.4 keV line strength is
easily consistent with reflection from a shock just above the white dwarf
surface i.e. with $\Omega/2\pi\sim 1$.

\section{CONCLUSIONS}

The BY Cam data from ASCA and GINGA give a physically self consistent picture of
an accretion column shock, cooling by both cyclotron and bremsstrahlung
emission. This multi--temperature X--ray continuum illuminates the white dwarf
surface, producing a reflection continuum and iron fluorescence line. Absorption
from the pre--shock column strongly modifies the observed spectral form, and is
a significant source of uncertainty since it depends on whether the pre--shock
column is uniform, or blobby (and hence ionised or nearly neutral), circular or
arc--like in cross--section and whether there is radial density structure. We
model this absorption by neutral material where the
covering fraction is a power law function of the column. This could represent a
physical situation where the column is blobby (unionised), with more blobs
accreting towards the centre of a circular column. However, it is more
likely that this form merely gives a suitable approximation to a more complex 
situation, and we urge further theoretical modelling of the pre--shock flow.

The multi--temperature emission and reflection model, together with the
power law neutral absorption model will be made publically 
available in the next release of XSPEC.

\section{ACKNOWLEDGMENTS}

We thank Dave Smith for his help with the GINGA data extraction, and Mark
Cropper for useful conversations and the use of his bremsstrahlung--cyclotron
cooling code.  CD acknowledges support from a PPARC Advanced Fellowship, and PM
acknowledges support from Polish Academy of Sciences and The Royal Society. This
research has been supported in part by the Polish KBN grant 2P03D01008 and has
made use of data obtained through the High Energy Astrophysics Science Archive
research Center Online Service, provided by the NASA/Goddard Space Flight
Center, and from the Leicester Database and Archive Service at the Department of
Physics and Astronomy, Leicester University, UK.

\end{document}